# Satellite-derived Land Surface Temperatures Strongly Mischaracterise Urban Heat Hazard

Wenfeng Zhan[a,b,c,1*], Benjamin Bechtel[d,1*], Huilin Du[a*], TC Chakraborty[e], Simone Kotthaus[f], E. Scott Krayenhoff[g], Alberto Martilli[h], Marzie Naserikia[i], Negin Nazarian[j], Matthias Roth[k], Panagiotis Sismanidis[d,l], Iain D. Stewart[m], and James Voogt[n*]

[a]*Jiangsu Provincial Key Laboratory of Geographic Information Science and Technology, International Institute for Earth System Science, Nanjing University, Nanjing, Jiangsu, China*
[b]*Jiangsu Center for Collaborative Innovation in Geographical Information Resource Development and Application, Nanjing, Jiangsu, China*
[c]*Frontiers Science Center for Critical Earth Material Cycling, Nanjing University, Nanjing, Jiangsu, China*
[d]*Bochum Urban Climate Lab, Ruhr-University Bochum, Bochum, Germany*
[e]*Atmospheric, Climate, and Earth Sciences Division, Pacific Northwest National Laboratory, Richland, WA, United States of America*
[f]*Laboratoire de Météorologie Dynamique, LMD-IPSL, École Polytechnique, Palaiseau, France*
[g]*School of Environmental Sciences, University of Guelph, Guelph, Ontario, Canada*
[h]*Centre of Research in Energy, Environment, and Technology (CIEMAT), Madrid, Spain*
[i]*Climate Change Research Centre, ARC Centre of Excellence for Climate Extremes University of New South Wales, Sydney, New South Wales, Australia*
[j]*School of Built Environment, ARC Centre of Excellence for 21st Century Weather University of New South Wales, Sydney, New South Wales, Australia*
[k]*Department of Geography, National University of Singapore, Singapore*
[l]*Institute for Astronomy, Astrophysics, Space Applications and Remote Sensing National Observatory of Athens, Athens, Greece*
[m]*Department of Geography, The University of British Columbia, Vancouver, British Columbia, Canada*
[n]*Department of Geography and Environment, Western University, London, Ontario, Canada*

[1]Wenfeng Zhan and Benjamin Bechtel contributed equally
***Corresponding authors**: Wenfeng Zhan, Benjamin Bechtel, Huilin Du, and James Voogt

**E-mail Addresses**: zhanwenfeng@nju.edu.cn (W. Zhan), benjamin.bechtel@rub.de (B. Bechtel), duhuilin_nju@foxmail.com (H. Du), tc.chakraborty@pnnl.gov (T.C. Chakraborty), simone.kotthaus@lmd.ipsl.fr (S. Kotthaus), skrayenh@uoguelph.ca (E. S. Krayenhoff), alberto.martilli@ciemat.es (A. Martilli), m.naserikia@unsw.edu.au (M. Naserikia), n.nazarian@unsw.edu.au (N. Nazarian), geomr@nus.edu.sg (M. Roth), panagiotis.sismanidis@rub.de (P. Sismanidis), iain.stewart@ubc.ca (I. D. Stewart), and javoogt@uwo.ca (J. Voogt)

**Abstract**
Escalating urban heat, driven by the convergence of global warming and rapid urbanization, is a profound threat to billions of city dwellers. The science directing urban heat adaptation is strongly influenced by studies that use satellite-based land surface temperature (LST), which is readily available globally and address data gaps in cities, particularly in the Global South. LST, however, is a poor surrogate for near-surface air temperature, physiologically relevant human thermal comfort, or direct human heat exposure. This flawed practice leads to issues for several downstream use cases by inflating adaptation benefits, distorting the magnitude and variability of urban heat signals across scales, and thus misguiding urban adaptation policy. We argue that satellite-based LST must be treated as a distinct indicator of surface climate, which, though relevant to the urban surface energy budget, can be frequently decoupled from human-relevant thermal impacts especially during daytime. Only by a disciplined application of this variable, combined with complementary datasets, process-based and data-driven models, as well as interdisciplinary collaboration, can urban adaptation design and policy be effectively advanced.

## I. Urban Heat Mitigation and Climate Adaptation in a Rapidly Urbanising World

The rising threat of urban overheating represents a defining challenge for humanity in the 21st century (Kalnay & Cai, 2003; Zhao, 2018; United Nations, 2019; Lin et al., 2021). Defined as a state where global climate change and local urban warming together drive temperatures beyond critical thresholds, this phenomenon has far-reaching and detrimental consequences for human health, comfort, and urban systems (Tuholske et al., 2021; Nazarian et al., 2022; Qian et al., 2022). The effects are not abstract: urban heat already amplifies mortality during heatwaves (Mora et al., 2017; Ballester et al., 2023), reduces labour productivity (He et al., 2022; Han et al., 2024), and undermines infrastructure resilience (Nazarian et al. 2022; Amonkar et al., 2023). Urban heat is also not evenly distributed within a city given the high levels of spatial heterogeneity, meaning communities can be exposed to different heat levels within the same city (Chakraborty et al., 2023; Rocha et al., 2024). In response to these challenges, cities worldwide are investing substantial sums in climate adaptation projects – from tree planting and cool roof implementation to heat action plans and redesigned public spaces (Estrada et al., 2017), including targeted strategies that can benefit vulnerable populations with less ability to adapt to these thermal stressors (McDonald et al., 2024).

However, effective adaptation requires reliable metrics. Decision-makers need to know which neighbourhoods are most exposed, who is most vulnerable at what times and during which season, and whether interventions are delivering measurable benefits (Reid et al., 2009). Assessment and monitoring are therefore not optional add-ons; they are the foundation for directing resources where they matter most. In this context, the choice of urban heat metrics is crucial (Patel et al., 2025). If the wrong variables are used to characterise urban overheating risk, adaptation strategies may become ineffective, inequitable, or even counter-productive (Vanos et al., 2020).

## II. Traditional Gaps and the Surge in Satellite-based Assessments of 'Urban Heat'

Before assessing the limitations of satellite-based assessments of 'urban heat' assessments, it is critical to elucidate what is being measured. From a thermodynamic perspective, 'heat' refers to the internal energy state of a system or object, which is generally proportional to temperature in the macro sense. Within the urban canopy, there are multiple objects and systems whose individual or combined temperatures can be captured in multiple ways. Classically, traditional in-situ approaches – whether using fixed meteorological stations or mobile transects – which directly record 2-meter screen-level surface air temperature (SAT), have predominated. SAT directly relates to the atmosphere's energy state and strongly influences the efficiency of its heat exchange with humans. As such, SAT is one determinant of pedestrian heat exposure. Beyond SAT, other meteorological variables, such as radiation (e.g., shade), humidity, and wind, also influence the net energy exchange between humans and their environment (Fan & McColl, 2024; Fig. 1). These drivers can be combined into other more comprehensive heat stress indices – including moist heat metrics such as heat index (HI), mean radiant temperature (Tmrt), and the universal thermal climate index (UTCI) – that better determine outdoor personal heat exposure and heat stress (Fig. 1).

Traditional in-situ observations offer critical insights into 'urban heat' but are inherently limited in scope. Station networks are often sparse and unevenly distributed, while mobile transects capture only localized gradients for a snapshot in time (Oke et al., 2017; Yang et al., 2024). The recent proliferation of crowdsourced networks has offered an adequate number of observations in certain cities (Meier et al., 2017). However, their coverage remains heavily confined to North America, Europe, and Australia, with only a handful of key metropolitan areas represented in Asia (Potgieter et al., 2021; Venter et al., 2021). This leaves significant gaps in the Global South, where observations are often sparse or entirely absent (Middel et al., 2022), hindering a truly global understanding of urban heat and its impacts.

Satellite thermal remote sensing has partly bridged such gaps. Since the 1970s, satellite-based land surface temperature (LST) has become central to urban heat and climate assessments and applications (Roth et al. 1989; Voogt & Oke, 2003; Zhou et al., 2018). This 'heat' is a bulk measure of the radiative temperature of the surfaces seen by remote sensors (Fig. 1). This is generally a lot more complicated and less physiologically-relevant than the 'heat' measured by ground sensors, as discussed in the next subsection. The past decade has witnessed an explosion of global analyses and local assessments that map 'urban heat' using satellite-based LST (Chakraborty & Lee, 2019; Roberts et al. 2023, New York City Interactive heat vulnerability index, 2025). These data are appealing – globally available, spatially detailed, and accessible from widely used platforms and sensors such as Landsat, Moderate Resolution Imaging Spectroradiometer (MODIS), ECOsystem Spaceborne Thermal Radiometer Experiment on Space Station (ECOSTRESS), and Sustainable Development Science Satellite 1 (SDGSAT-1; Guo et al., 2023). Researchers and practitioners alike are tempted to use them as a shortcut to fill gaps where dense SAT and human thermal comfort monitoring are lacking (Zhou et al., 2018; Diem et al., 2024). Indeed, LST-based analyses are increasingly cited in policy documents, municipal risk assessments, and even international adaptation funding proposals. Global reports highlight surface hotspots, and city governments overlay satellite images to identify priority neighbourhoods for tree planting or cooling centres. Easily converted into intuitive maps, at first sight satellite LST products offer a compelling visual narrative of urban heat (Fig. 1).

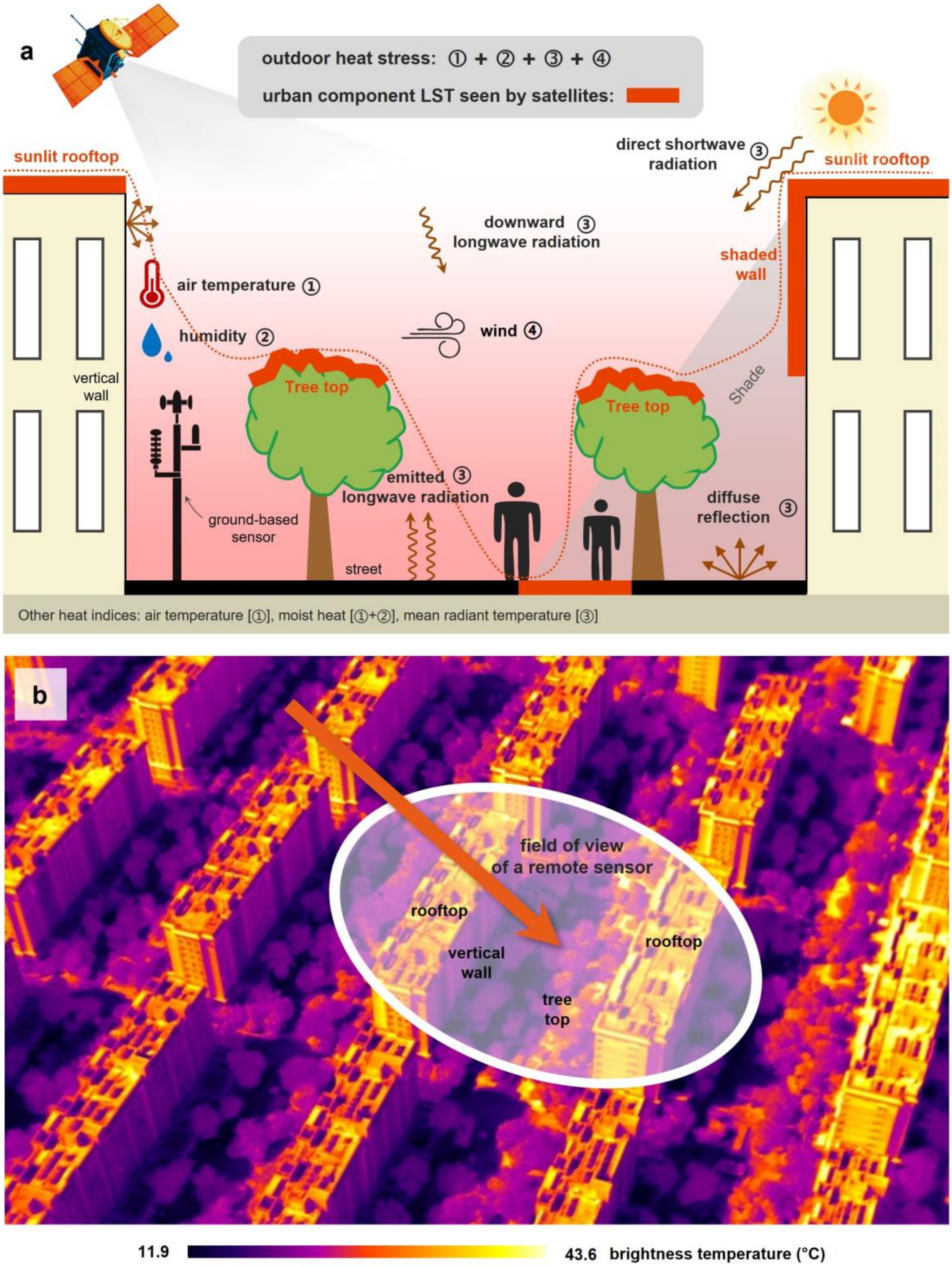

**Fig. 1. Outdoor heat stress versus urban land surface temperatures (LSTs) measured by remote sensing | (a)** In a typical urban canyon, outdoor heat stress is shaped by surface air temperature, humidity, radiation (shortwave and longwave), and wind speed. Indices commonly used include

SAT, moist heat indices (integrating SAT and humidity), mean radiant temperature (Tmrt), and comprehensive metrics such as the universal thermal climate index (UTCI). By contrast, remote sensors primarily measure longwave radiation emitted from the outermost surfaces within their field of view (orange dashed line), principally integrating thermal emissions from rooftops, tree-tops, and a portion of building walls and pavements. **(b)** A thermal image (brightness surface temperatures) captured by a DJI drone on 23 September 2021 at 10:15 AM over a residential area (in Nanjing, China) illustrates the urban surface types typically measured by remote sensors.

This allure, however, obscures a critical problem: LST substantially differs from both SAT and personal heat exposure (Parlow, 2021; Kelly Turner et al., 2022; Li et al., 2023). Unlike traditional in-situ air temperature, LST denotes a directional component of the heat emitted as longwave radiation from the Earth's outermost surface layer, which is then expressed as a 'radiative temperature' (Norman & Becker, 1995). Many researchers and practitioners – particularly those outside of thermal remote sensing and urban climatology – lack a clear grasp of what LST actually represents, how it differs from SAT and thermal comfort, and its inherent limitations. As a result, using LST directly as a proxy for human heat stress or to monitor the effectiveness of heat adaptation strategies can be deeply misleading. Such misapplications are costly as cities attempt to deal with intensifying overheating. Understanding the nature and limitations of LST is essential to avoid its potential to misguide future urban heat adaptation efforts.

## III. <u>What Satellite LST Really Measures and Why it differs from SAT and heat stress</u>

Satellite LST primarily denotes the longwave thermal emission from surfaces that have a large exposure to the sky within a sensor's field of view. The measured signal represents an integrated footprint (source area) of rooftops, the tops of vegetation canopies, parts of building walls, and street surfaces (Voogt & Oke, 2003). These measurements are made from a limited selection of viewing positions – most often directly downwards but some from an oblique angle – and thus instantaneous snapshots of the surface temperature at the acquisition time, which is usually limited to regular satellite overpasses (Figs. 1 & 2). The measurements therefore hide important microscale details of urban surface temperature differences that are relevant to applications and are inherently anisotropic – meaning the measured temperature depends on the viewing direction (Du et al., 2023).

Four key dimensions define the nature of satellite-based LST – temporal snapshots, spatial mixing, angular bias, retrieval uncertainty (Fig. 2). The combination of these four key dimensions means that there are various artefacts in LST that are functions of the satellite overpass and observation geometry, its spatiotemporal resolution, and retrieval algorithm, showing different magnitude of urban surface 'heat' (Botje et al., 2022).

**<u>Temporal snapshots</u>**: LST is captured only at the moment of a satellite overpass. Constrained by satellite orbital cycles and revisit intervals, these observations may occur just once or twice daily – or even weeks apart – missing variation over the full daily cycle (Fig. 2c). Clouds are an additional limiting factor, with LST being completely unavailable under cloud cover. Due to temporal snapshots, the currently available polar-orbiting satellites (e.g., Landsat) often miss hourly dynamics within a diurnal cycle, most critically the nocturnal heat retention that drives excess mortality during heatwaves (He et al., 2022). Furthermore, cloud cover can systematically bias LST retrievals because missing data often coincide with periods of extreme heat. This leads to an underrepresentation, and sometimes even misrepresentation, of when and where urban heat stress is most pronounced. This challenge is especially severe in tropical regions where heat stress is a leading driver of morbidity and mortality (Diem et al., 2024).

**Spatial mixing**: Each LST pixel integrates thermal radiance from a heterogeneous mix of surfaces – rooftops, walls, pavements, water, and vegetation – each having a different surface temperature. It also incorporates both radiant emissions from these surfaces that depend on the surface temperature and reflected radiances from the surfaces which depend on the respective emissivity, a largely temperature-independent radiative property of the surface. Furthermore, satellite LST represents a biased spatial sample of surface temperature. It does not register the full active surface of urban terrain, but tends to over-emphasize the role of roofs and tree-tops while under-sampling vertical surfaces and surfaces underneath tree canopies (Voogt & Oke, 1997) – precisely the micro-environments most critical for outdoor human heat stress and its potential relief (Fig. 1). Further, instruments with moderate and coarse spatial resolution mix the signal from ground surfaces with those from roof and tree tops, providing spatial mixing of the viewed surfaces into a single pixel value. Depending upon the urban geometry, the pixel value may be disproportionately weighted in favor of one surface type (the dominant one or the one with the highest surface temperature), and therefore represent a location which has little effect on the urban pedestrian-level climate (Roth et al., 1989).

**Angular bias**: The complex three-dimensional urban morphology generates patterns of illumination and shadow. Since urban surfaces differ widely in surface temperature, varying proportions of these surfaces within the sensor's field of view can shift the measured LST with the viewing angle a phenomenon known as the effective urban thermal anisotropy effect (Voogt, 2008; Hu et al., 2016). If unrecognized, satellite LST differences across angles may be misinterpreted as genuine spatial variations in 'urban heat', even within the same area, when they are in fact an artifact of observation geometry. Beyond misinterpretation, this effect further introduces systemic biases when directly comparing across satellites, or even across repeated observations by the same satellite, therefore posing a major obstacle to assess the true spatiotemporal performance of cooling strategies.

**Retrieval uncertainty**: Accurate LST retrieval necessitates precise inputs of emissivity estimates and atmospheric characteristics. However, emissivity estimates are ill-defined for mixed pixels incorporating different urban materials (Norman & Becker, 1995), and are often prescribed using various algorithms that are distinct for different satellite products (Chakraborty et al., 2021). Many available LST products rely on empirical emissivity values calibrated for natural surfaces, not cities (Sobrino et al., 2004; Wan, 2008), where the range of building materials can lead to significant variabilities in this critical input to LST retrieval algorithms (Chen et al., 2016). Furthermore, most operational satellite LST products disregard intra-urban variability or even differences between urban areas and their surroundings with respect to atmospheric characteristics, such as aerosols, clouds, and water vapor (Vo et al., 2023). The added complexity of atmospheric characteristics also precludes accurate atmospheric correction and subsequent LST retrieval. These retrieval uncertainties, together with those induced by urban thermal anisotropy, can introduce significant uncertainties: over urban surfaces, LST errors often exceed nominal dataset accuracies, reaching 3.0 to 5.0 °C or more (Li et al., 2023). Such errors can rival the very warming signals they are intended to detect, making attribution to climate change or to specific adaptation interventions especially challenging and in some cases impossible.

By comparison, SAT is usually defined as the 2-meter screen-level air temperature measured within street canyons, on open courtyards, or beneath vegetation canopies. SAT reflects the combined influence of all urban surface energy fluxes and is an integrated signal of urban land–atmosphere interactions. Moreover, it is strongly affected by atmospheric motion (advection and turbulence) – while local exchanges strongly affect SAT under very calm conditions, mixing and advection of air from upwind areas can influence SAT when winds are present (Fig. 2).

Therefore, LST and SAT differ sharply in both temporal and spatial patterns (Figs. 2 & 3). During the day, especially under clear skies, LST can exceed SAT by 20 to 30 °C for some individual surfaces, while at night LST often falls slightly below SAT (Fig. 2c). While seasonal and inter-annual trends generally align, spatial and temporal variations in LST are far more pronounced than those in SAT (Naserikia et al., 2024; Du et al., 2025). Within cities, LST and SAT show stronger coherence at night, yet daytime spatial correlations are weak and sometimes absent and the diurnal cycle of the Urban Heat Island (UHI) is usually opposite for LST and SAT (Fig. 2c) due to differences in the physical mechanisms controlling the two variables (Venter et al., 2021; Krelaus et al., 2024). Furthermore, while one might expect qualitatively similar results for the spatial variability of LST and SAT for sprawling cities when aggregated to larger neighborhoods, the magnitudes are still considerably overestimated when using clear-sky LST (Chakraborty et al., 2023; Du et al., 2025). Even high-resolution satellite LST products cannot resolve this disconnect during daytime, as the underlying physics of surface heating and air mixing are fundamentally different (Fig. 4). The relationship between LST and SAT also shifts with time-of-day, season, urban geometry, and material properties, with the strongest decoupling on hot, dry days (Venter et al., 2021; Naserikia et al., 2024).

Likewise, the relationship between LST and heat stress indicators can be weak. For instance, for cities with significant vertical surfaces, where shadows can impact LST estimates, LST can be completely decoupled from more comprehensive heat stress metrics (Fahy et al., 2025). This also applies to moist heat metrics that integrate SAT and humidity (e.g., the HI). For example, the spatial patterns of the HI also do not align with LST-defined hotspots and cool spots within cities (Fig. 4b).

Moreover, in cities with dense vertical structures, shading can substantially alter Tmrt, producing distinctly different spatial and diurnal patterns relative to those in unshaded areas where values peak in the afternoon and drop at sunset (Kelly Turner et al., 2022; Fahy et al., 2025).

In summary, satellite LST provides valuable information on the surface energy state of the sensed surfaces, but is not a direct quantitative proxy for SAT or heat stress. These distinctions are critically important when applying satellite LST to urban health, planning, and design.

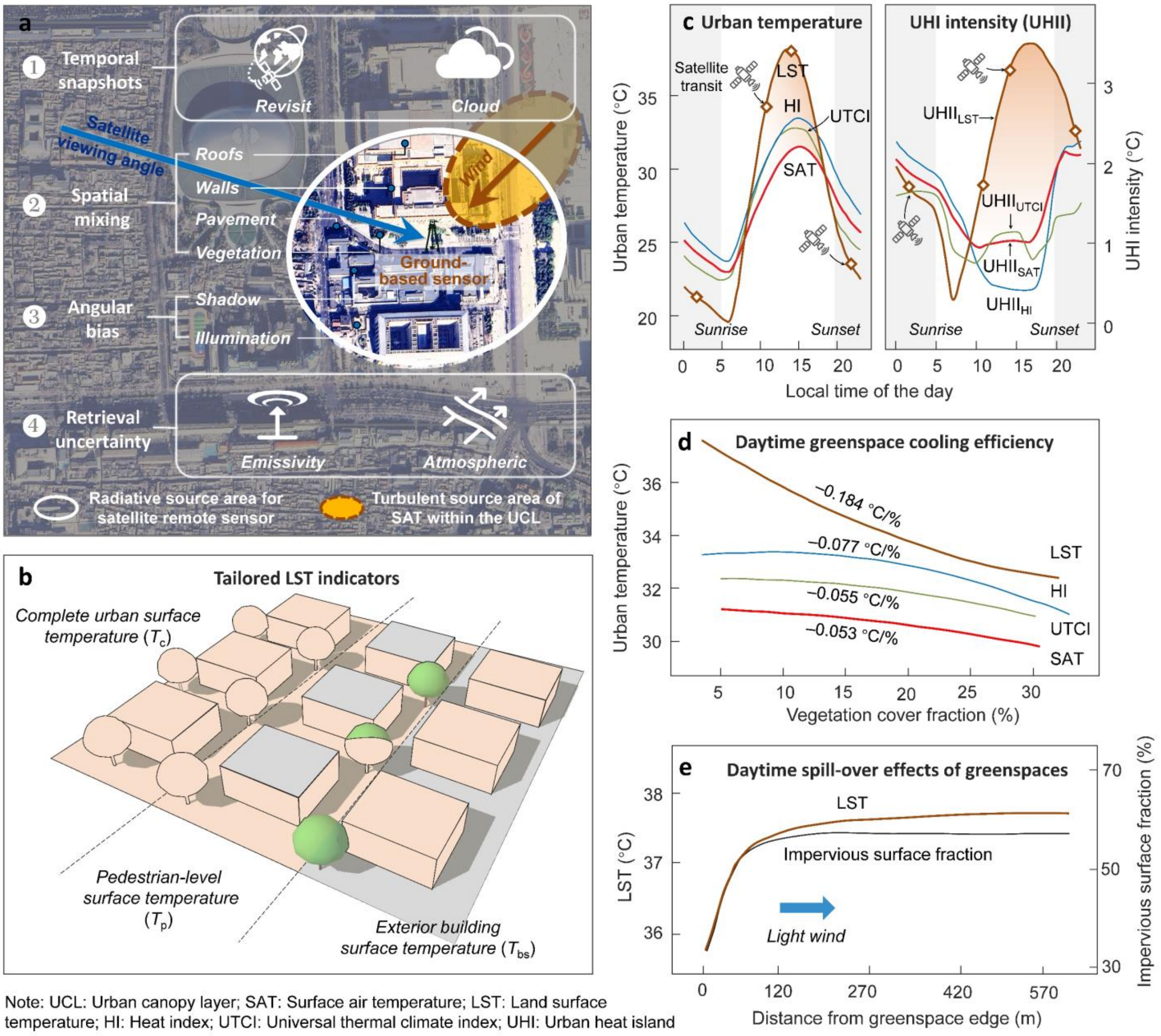

**Fig. 2. Characterizing satellite-based land surface temperatures (LSTs) and their applications in urban heat studies |** (**a**) The difference in source area between surface air temperature (SAT) and satellite LST, as well as the four key dimensions underlying what satellite LST really represents: temporal snapshots, spatial mixing, angular bias, and retrieval uncertainty. (**b**) Application-specific urban LST indicators more directly linked to urban processes, including complete urban surface

temperature ($T_c$) for surface-atmosphere interactions, pedestrian-level surface temperature ($T_p$) for population heat exposure, and exterior building surface temperature ($T_{bs}$) for building energy budgets (Stewart et al., 2021). Surfaces marked in light orange are included in the application-specific urban LST indicators, while gray surfaces are not. (**c**) Diurnal variations of various urban temperature metrics, including LST, SAT, heat index (HI, based on SAT and humidity), and the Universal Thermal Climate Index (UTCI, based on SAT, humidity, wind, and radiation), along with their corresponding urban heat island (UHI) intensity estimates in a typical temperate city (Beijing, China) during summer (averaged over the period 2017 to 2019), calculated based on a meteorological station network with 58 urban sites and 15 rural ones. Hourly meteorological data under all-weather conditions were obtained from the China Meteorological Data Service Centre (Wang et al., 2022), while hourly LST was reconstructed by integrating four MODIS overpasses under clear-sky conditions with a previously developed four-parameter diurnal temperature cycle model (Hong et al., 2018). (**d**) Greenspace cooling efficiency as assessed by different urban temperature metrics in Beijing during summer daytime, also based on the same meteorological station network as used for (**c**). (**e**) Radial gradients of LST and impervious surface fractions (ISF) outward from 701 urban parks larger than 5.0 hectares in Beijing, illustrating the daytime cooling (i.e., spillover of cooling) effects of greenspace.

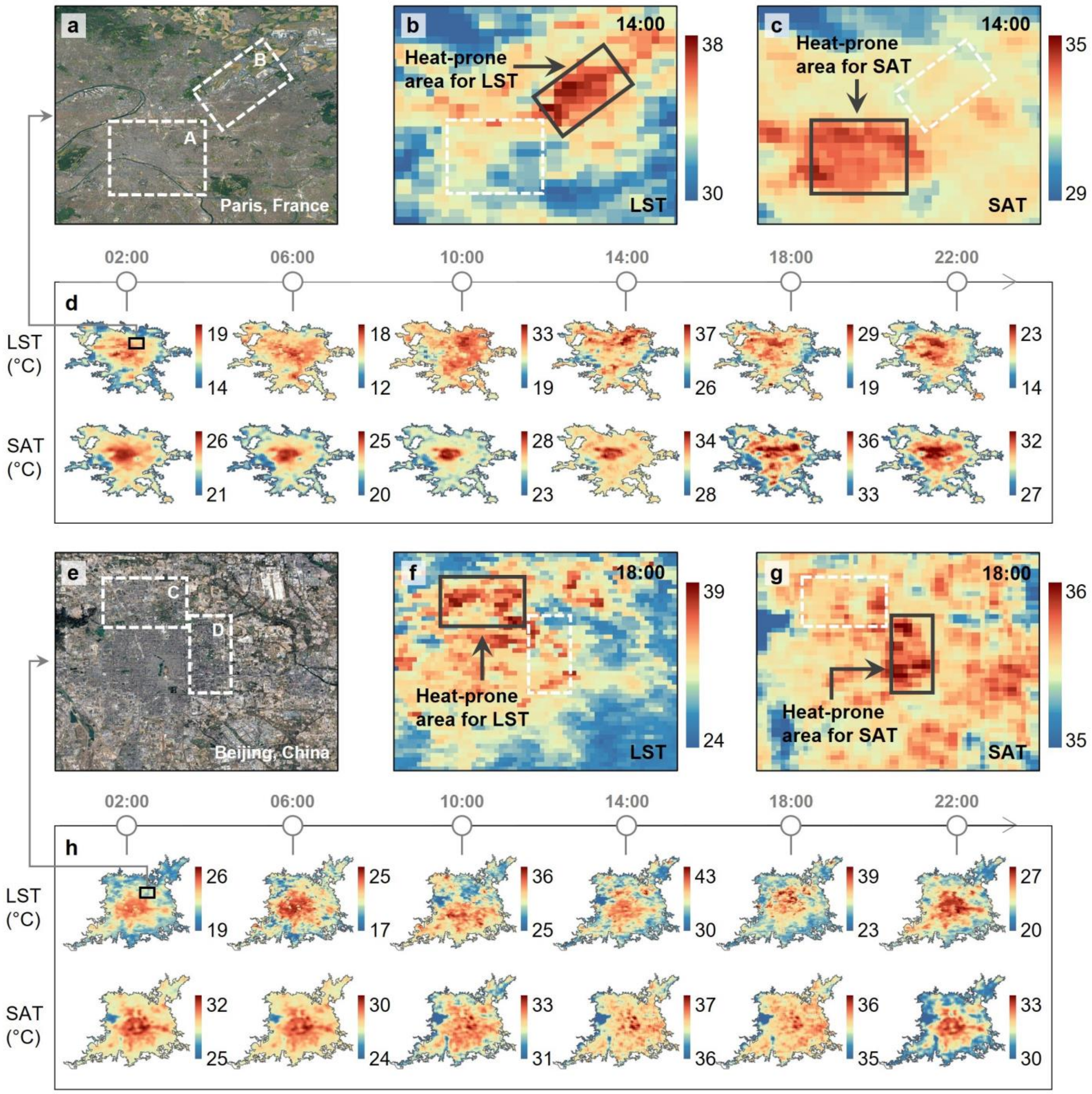

**Fig. 3. Contrasting spatial patterns of land surface temperature (LST) and surface air temperature (SAT) in two temperate cities during a diurnal cycle |** (**a**) Google imagery of Paris, France, highlights the urban core (box A) and a northeastern suburb (box B). (**b**) Monthly mean 1-km resolution LST at 14:00 local time in August, 2024, underlines a hotspot zone in box B. (**c**) The corresponding 1-km resolution SAT during the same time period identifies a high-temperature zone in box A. (**d**) Intra-day spatial patterns of LST and SAT across Paris are displayed (i.e., the panels at the third and fourth rows). (**e**) Google imagery of Beijing, China, shows a dense residential zone (box C) and a transitional area (box D). (**f**) Monthly mean 1-km resolution LST at 18:00 local time in August, 2018, reveals a hotspot zone in box C. (**g**) The corresponding 1-km resolution SAT during the same time period, yet indicates a high-temperature zone in box D. (**h**) Intra-day spatial patterns

of LST and SAT across Beijing are displayed (i.e., the panels at the fifth and sixth rows). Note that hourly LSTs were derived by combining MODIS LST products and a diurnal temperature cycle model (Hong et al., 2018). For SAT, estimates in Paris combined more than 3,000 crowdsourced stations with a well validated deep learning approach, whereas those in Beijing used over 100 urban stations with the same approach (Ge et al., 2025).

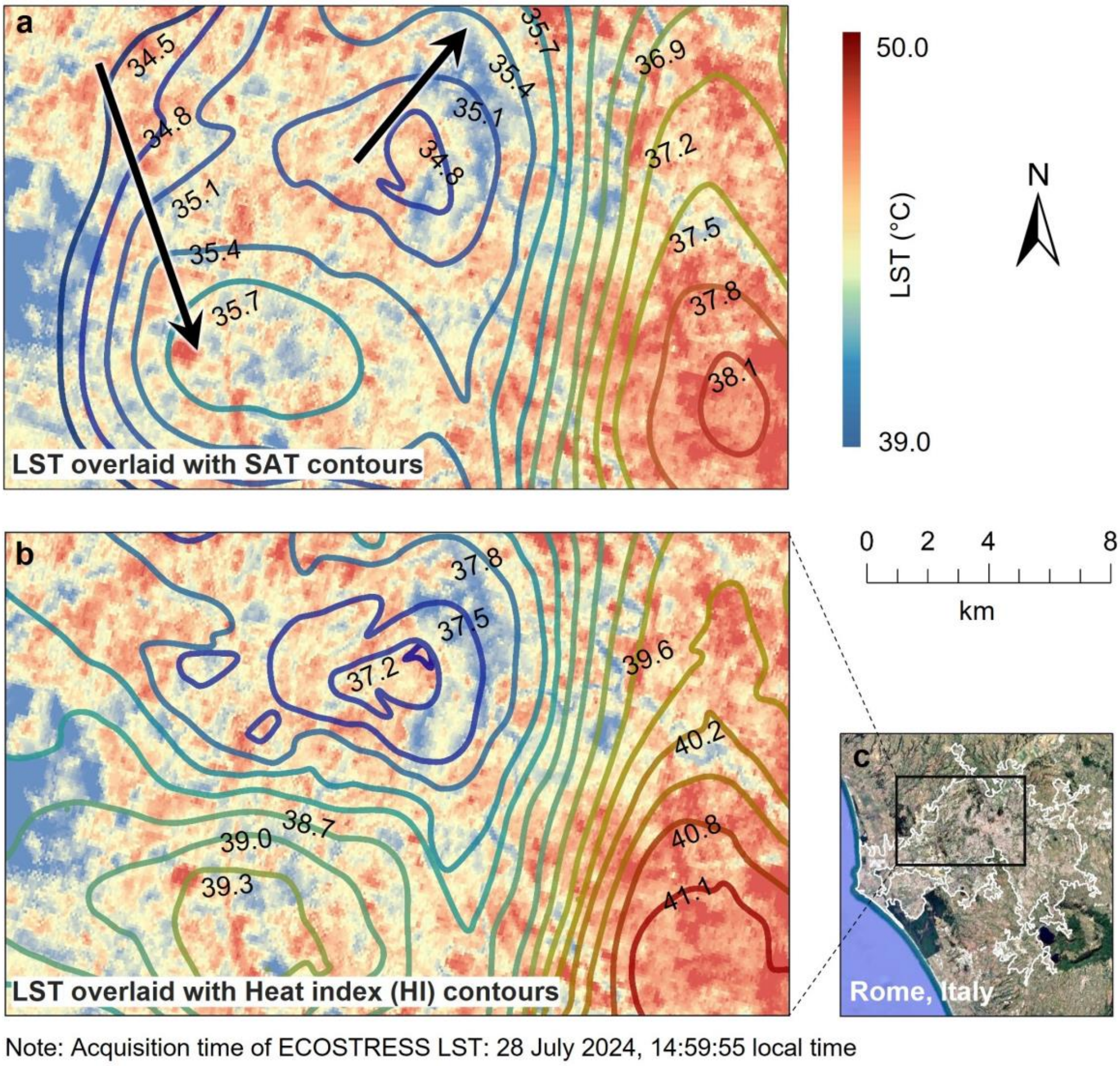

**Fig. 4. Spatial contrasts among LST, SAT, and HI in the urban core of Rome, Italy|** (**a**) 70-m ECOSTRESS LST overlaid with SAT contours. (**b**) LST overlaid with HI contours. LST was acquired on 28 July 2024 at 14:50 local time. SAT and HI contours (0.3°C intervals) were interpolated from 195 crowdsourced monitoring stations within this region. In (**a**), black arrows indicate two areas where LST declines while SAT rises along the arrow direction.

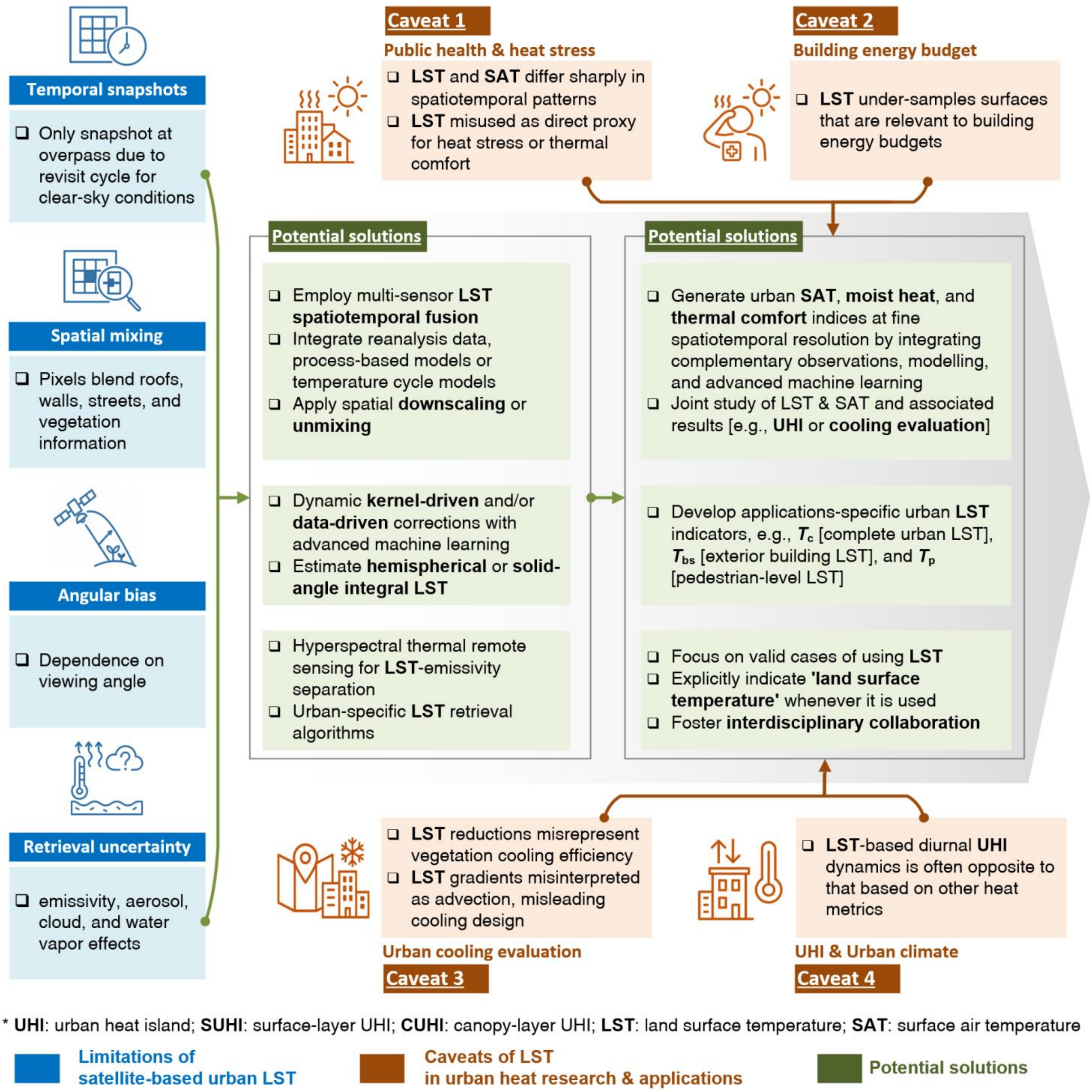

**Fig. 5. Limitations and misapplications of satellite-derived land surface temperature (LST) in urban heat research and potential solutions.**

## IV. <u>Challenges and Caveats in Applying Satellite-based Urban Heat Estimates</u>

Because LST interacts with the atmosphere and urban surfaces in highly complex ways and represents a biased spatial sample, its relationship to SAT, and thermal comfort is both nonlinear and indirect (Fig. 2). Glossing over this fundamental distinction has led to recurring issues in several areas of urban research and policy as detailed below.

- **Heat stress and thermal comfort**

**Weak linkage to urban heat stress**: For heat stress assessment, the disparity both in magnitude and spatial variation between LST and SAT is not incidental, but systematic and consequential. When considering health risks, the mismatch grows wider: outdoor thermal comfort depends not only on SAT but also on humidity, radiation, and wind (Blazejczyk et al., 2012) – factors mostly unrepresented by LST. Physiologically-relevant radiation offers a clear case: daytime outdoor thermal comfort is well linked to shadow-driven radiation patterns under tree canopies within urban canyons, which is difficult to capture with satellite LST (Gu & Zhang, 2025). Nighttime correlations between satellite LST and outdoor thermal comfort are somewhat stronger, but critical contributions emitted by streets, building walls, and trees are often missing or poorly represented in satellite LST (Roth et al., 1989). In fact, especially during daytime, many urban 'heat' signals that appear strong in LST maps become substantially muted for SAT and especially for moist heat stress indices (Chakraborty et al., 2022). Relying on LST alone to infer, especially quantify, outdoor heat stress is deeply misleading – and potentially dangerous when performing downstream cost-benefit analyses for informing the allocation of critical public health resources.

- **Building heat budgets and design**

**Methodological constraints in building studies**: Although buildings are central to urban heat exposure and energy use, LST poorly represents their surface thermal dynamics (Fig. 2b). Building energy consumption for space cooling (or indoor heat stress), cannot be assessed without considering urban form and building thermal properties. A building with roof top insulation may have higher roof top surface temperature, but lower building energy consumption than a building of the same size without insulation (Piselli et al., 2019). Coarse pixels (>100 m) integrate thermal signals from a heterogeneous mix of roofs, streets, vegetation and sometimes walls (Stewart et al., 2021). Nadir-viewing sensors (e.g., Landsat TIRS) primarily capture mostly roof, tree-tops, and open areas, while wide-swath sensors with off-nadir viewing capability (e.g., MODIS) include some walls but still yield mixed signals and many surfaces are occluded as the view angle away from nadir increases. As a result, using LST directly for building surface energy budgets or design guidance can yield misleading conclusions about energy efficiency or the quantitative performance of specific architectural forms.

- **Urban cooling evaluation**

**Inconsistent cooling capacity estimates at local scales**: Vegetation is a widely promoted cooling strategy, but assessments based on spatial variations in LST alone may misrepresent its true effect (Schwaab et al., 2021). Because evapotranspiration strongly suppresses daytime LST, green spaces tend to have obviously lower surface temperatures than their built-up surroundings (Fig. 2d), while effects on SAT are generally more modest (Du et al., 2024). In fact, neither LST nor SAT adequately captures the daytime cooling effects from tall vegetation that are largely linked to their ability to shade the underlying ground or nearby building and pedestrians. Clearly, spatial contrasts in LST between tree-tops and impervious surfaces cannot be interpreted as suitable indicators for daytime heat mitigation potential. Moreover, LST analyses often suggest a reduced cooling efficiency – measured as temperature reduction per unit vegetation – when vegetation cover increases (Zhou et al., 2021), whereas SAT studies suggest that cooling efficiency may strengthen with more vegetation (Fig. 2d; Ziter et al., 2019). Relying solely on LST risks an inadequate assessment of the benefits of urban green infrastructure and misdirecting cooling investments. Even worse, reflective cool pavements increase albedo and decrease LST and thus the emitted longwave radiation, yet at the same time they intensify shortwave radiation reflection toward pedestrians and buildings, and therefore daytime heat stress (Schneider et al. 2023). Finally, indicators of urban heat and mitigation efficiency risk being misleading if they fail to differentiate between daytime and nighttime processes: Solar insolation and atmospheric mixing govern daytime heat exchange, whereas thermal radiation exchanges play a considerably more important role at night, with considerably different consequences for both human thermal comfort and building energy budgets (Haeffelin et al., 2024).

**Misinterpreted cooling capacity estimates at nonlocal scales**: Hundreds of studies have used satellite LST to estimate the "spill-over effects" of green spaces, which are presumed to cool nearby areas through thermal advection (Galalizadeh et al., 2024). However, LST gradients extending outward from parks, align well with impervious surface fractions (Fig. 2e), suggesting these patterns primarily replicate surface composition rather than air circulation. Studies relying exclusively on LST to infer that specific urban land cover types cool adjacent or distant areas via thermal advection should be interpreted cautiously. LST data alone cannot disentangle whether observed temperature reductions result from atmospheric circulation or variations in surface properties. Designing urban cooling strategies on LST contrasts alone risks both misleading attribution and misguided interventions.

- **UHI dynamics and urban climate**

**Divergent diurnal UHI dynamics across temperature metrics**: Surface and canopy UHIs (SUHI and CUHI) diverge strongly in their diurnal cycles (Fig. 2c). In most non-arid cities, the LST-based SUHI intensifies during the day and weakens at night, while the SAT-based CUHI typically peaks a few hours after sunset and diminishes – or even vanishes – during daytime (Lai et al., 2018; Stewart et al., 2021; Venter et al., 2021). These contrasts underscore that UHIs derived from LST and those based on SAT or human-relevant indices are fundamentally non-interchangeable when used to compare the local urban thermal impacts on humans over the diurnal cycle.

## V. Research and Policy Pathways Forward

Recognizing these limitations and caveats does not mean discarding satellite LST entirely. It remains valuable for advancing our understanding of urban climates (Weng, 2009). For instance, at aggregated scales, city-average LST trends are valuable indicators for long-term trend analysis of urban thermal environmental change (Naserikia et al., 2024), provided they are not misinterpreted as quantitative proxies for changes in heat exposure or risk (Liu et al., 2022; Shen et al., 2023). Furthermore, satellite LST, when combined with other remotely sensed biophysical indicators – such as vegetation indices – can be used to derive urban surface energy fluxes that are essential for understanding urban boundary-layer processes and climate (Chrysoulakis et al., 2018). Here we advocate for a more disciplined application of LST, leveraging its strengths in combination with complementary data and models to build a robust foundation for urban adaptation. We further advocate the following steps forward (Fig. 5).

**Explicitly label as LST-based findings in visible places**: Whenever LST is employed to study urban climates, UHIs, or warming, the findings must be explicitly presented as indicative of the urban 'surface' climate – specifically, 'surface' UHI or 'surface' warming (Fig. 5). Furthermore, publications must clearly denote in titles and abstracts that results are LST-based findings, rather than ambiguously labelling them as 'temperature'. For example, when evaluating cooling effects, one should explicitly specify 'LST reduction' instead of the vague term 'cooling'; if 'cooling' must be used, it should be written as 'LST-based cooling'. Without this simple but critical clarity, results risk being misinterpreted as representing air temperature or direct heat exposure. This practice would prevent cross-disciplinary misunderstandings and ensure that findings arising from using LST are correctly contextualized. Moreover, studies using LST should avoid framing their results as indicative of a heat hazard or exposure. Critically, researchers should also delineate what LST represents, and what it does not, in visible places of their publications: LST-based findings characterize surface radiative properties, not direct metrics of human heat stress. Communicating this distinction is essential for cross-disciplinary users – such as urban planners, humanities and social scientists, and even the general public – who may otherwise misinterpret 'cooling' as a direct reduction in human heat exposure.

**Improve LST products**: Urban LST datasets must deliver finer spatial resolution, more seamless temporal coverage, and resolve thermal anisotropy (Fig. 5). Achieving this requires integrating multi-sensor spaceborne, airborne, and unmanned aerial vehicle (UAV) observations − including microwave data less affected by clouds − with reanalysis products and mesoscale process-based models to close spatiotemporal gaps (Zhang et al., 2021; Chen et al., 2024). The latest geostationary meteorological satellites, including Japan's Himawari and the U.S. GOES-R series, now provide LST at less than 15-minute intervals and 2-km resolution (Schmit et al., 2017). Recent polar-orbiting missions from China enhance this detail: SDGSAT-1 (30 m) and ZY1-02E (15 m, down to 8 m with super-resolution) offer unprecedented spatial fidelity (Ouyang et al., 2024; Wang et al., 2025), while the forthcoming Land Surface Temperature Monitoring (LSTM) and Thermal Infra-Red Imaging Satellite for High-resolution Natural Resource Assessment (TRISHNA) missions, will provide nighttime observation over cities and allow the simultaneous retrieval of the LST and surface emissivity. Collectively, these platforms or sensors, establish the foundation for next-generation urban LST datasets with improved spatiotemporal resolution and accuracy. Furthermore, supported by multi-source, multi-temporal, and multi-angle thermal observations, dynamic kernel-driven and/or data-driven approaches show promise in correcting directional effects and generating hemispherical radiative temperatures over heterogeneous urban surfaces (Wang et al., 2020; Qin et al., 2023; Du et al., 2025). In addition, new robust methods for quantifying the LST retrieval uncertainty have been developed in the new state-of-the-art LST data products developed by the European Space Agency Climate Change Initiative (Ghent et al., 2019). Without such advancements, the utility of LST for urban climate and health applications remain fundamentally limited.

**Develop more application-specific urban LST indicators**: Generic urban LST is inadequate for the complexity of urban systems given it represents an incomplete and biased sampling of the urban surface (Fig. 2). An appropriate LST indicator is one that matches the intended application, including metrics of complete urban surface temperature that better capture land–atmosphere interactions, pedestrian-level surface temperature which is more closely linked to human heat stress, and building surface temperature which is more appropriate for building energy budgets (Stewart et al., 2021). For this, it is critical to consider the different processes that drive urban heat during convective daytime conditions vs those at night with low turbulent mixing. Multi-angle thermal sensing can help estimate hemispherical radiative temperatures as proxies for complete urban surface temperatures (Chang et al., 2025) and be extended to estimates of pedestrian-level and building surface temperatures through optimized solid-angle integrals (Jiang et al., 2018). Combined with visible-thermal unmixing or downscaling (Hu et al., 2024), these approaches offer powerful routes to derive urban component surface temperatures that could be the basis for estimating application-specific surface temperatures. Emerging high-resolution satellite thermal sensors, together with the increasing availability of refined LST products, would also transform our ability to estimate urban surface component temperatures. These finer estimates open new opportunities for application-specific analyses (e.g., building energy budget modeling), especially when combined with detailed morphological information. Such application-relevant surface temperatures can better bridge urban climate science, health, and urban adaptation planning. Nevertheless, even these application-specific urban LST indicators omit key factors influencing pedestrian heat exposure or building heat load, such as pedestrian shading or building wind load, and must therefore be used with caution, and ideally with complementary data that contributes to more holistic assessment of heat exposure.

**Integrate LST with complementary observations and modelling to model human-centric heat:** A key priority is to produce thermal comfort metrics directly relevant to human health – such as SAT, HI, and the UTCI – at fine spatiotemporal resolution across global cities. Addressing this priority requires a combination of diverse and complementary data streams: satellite LST, other Earth observations, reanalysis products, and in-situ measurements obtained from fixed stations, crowdsourced networks, mobile platforms, and vertical profiles captured either directly or via ground-based atmospheric remote sensing (Haeffelin et al. 2024). The real promise lies in integrating these diverse data streams with process-based models and/or advanced deep learning methods (Fig. 5). In addition, algorithms and high-resolution products on human-centric heat metrics must be readily scalable across global cities. When paired with cloud computing for big-data processing (e.g., Google Earth Engine), these relevant algorithms hold substantial potential for global application (Li et al., 2025). Moreover, satellite LST can be used to evaluate mesoscale urban canopy models (Hall et al., 2024) and street-scale microclimate models to improve simulations of street-level thermal conditions. While such integrated models should be the default evidence base for climate adaptation planning, critical obstacles remain in developing standards for validation and operational deployment.

**Strengthen interdisciplinary dialogue:** Beyond technical integration, harnessing the full potential of LST demands bridging disciplinary divides among remote sensing experts, urban climatologists, public health experts, architects, and city planners (Fig. 5). This represents a cultural rather than a technical shift. Thermal remote sensing experts are generally aware of the inherent uncertainties and physical assumptions underlying LST data. However, while such caveats are routinely addressed in the remote sensing literature, their communication is often absent when LST is adopted by other fields for general-interest audiences, risking misinterpretation and therefore requiring sufficient clarification to avoid misleading readers. In parallel, urban practitioners, drawing on these interdisciplinary applications, must also recognize and account for these important uncertainties to ensure appropriate use of satellite-based LST. Only through co-production of knowledge can we ensure that satellite-based LST fulfils its potential for urban resilience (Grimmond et al., 2020), so as to avoid the allure of attractive but potentially misleading urban planning support.

## VI. Concluding Remarks

Satellite-based LST has fundamentally transformed our understanding of urban thermal landscapes, revealing unprecedented heterogeneity within cities worldwide. However, the urgency of effective urban heat adaptation demands metrics that are credible, quantitatively and qualitatively relevant, and scientifically defensible. We argue that satellite LST, though convenient and available, does not meet this standard when used in isolation. LST is not a universal proxy since it cannot be equated to air temperature, nor thermal comfort or direct urban heat exposure. Misusing LST risks incorrect estimates of adaptation benefits, misallocating critical resources, and ultimately misinforming evidence-based urban climate policies.

For scientists, the imperative is to focus on valid use cases, explicitly refer to 'surface temperature' whenever conclusions are based primarily on LST, refine urban LST retrievals, define more application-specific urban LST indicators, and generate fine-resolution urban air temperature and human-relevant heat metrics by incorporating complementary observations and modelling. Looking ahead, urban climate informatics – fusing dense sensor networks, non-traditional datasets, and advanced machine learning – promises to deliver more human-relevant heat services and decision support (Middel et al. 2022). For urban planners and policymakers, the challenge is to demand evidence that truly reflects actual human exposure, rather than seductive but misleading surface heat maps derived solely from satellites. Misplaced confidence in LST could inadvertently lead to distorted urban adaptation strategies at the very moment when they are most needed.

The way forward is not to abandon LST but to apply it with greater scientific rigor. Used alone, LST risks becoming a mirage – visually captivating but misleading. Used wisely, especially when integrated with complementary data and models, and interpreted through interdisciplinary collaboration, LST can remain an invaluable tool for building sustainable and resilient cities under a warming climate. Given that cities will have to invest heavily in their climate resilience, we cannot afford to 'cool pixels while people swelter'.